\documentclass[conference]{IEEEtran}
\usepackage{times}
\usepackage{color}
\usepackage[ruled, vlined]{algorithm2e}
\usepackage[pdftex]{graphicx}
\usepackage{amsmath}

\DeclareGraphicsExtensions{.pdf,.jpeg,.png,.eps}

\begin{document}
\title{Graphical Probabilistic Routing Model for OBS Networks with Realistic Traffic Scenario}

\author{Martin L\'evesque and Halima Elbiaze\\
Department of Computer Science\\
Universit\'e du Qu\'ebec \`a Montr\'eal\\
Montr\'eal (QC), Canada\\
Email: levesque.martin.6@courrier.uqam.ca, elbiaze.halima@uqam.ca}



\maketitle

%
\begin{abstract}

Burst contention is a well-known challenging problem in \textit{Optical Burst Switching} (OBS) networks.
Contention resolution approaches are always reactive and attempt to minimize the
BLR based on local information available at the core node. On the other hand,
a proactive approach that avoids burst losses before they occur is
desirable. To reduce the probability of burst contention, a more robust routing algorithm than the shortest path is needed. This paper proposes a new routing mechanism for JET-based OBS networks, called \textit{Graphical Probabilistic Routing Model} (GPRM) that selects less utilized links, on a hop-by-hop basis by using a bayesian network. We assume no wavelength
conversion and no buffering to be available at the core nodes of the OBS
network. We simulate the proposed approach under dynamic load to
demonstrate that it reduces the \textit{Burst Loss Ratio} (BLR) compared to static approaches by using \textit{Network Simulator 2} (ns-2) on NSFnet
network topology and with realistic traffic matrix. Simulation results clearly show that the proposed approach outperforms static approaches in terms of BLR.
\end{abstract}

\IEEEpeerreviewmaketitle


\section{Introduction}

\textit{Optical Burst Switching} (OBS) \cite{obsGeneral, obsGeneral2, gaugerThesis}
is a promising technology to handling bursty and
dynamic Internet Protocol traffic in optical networks effectively. In OBS networks, user data (IP for example) is assembled as a huge
segment called a \textit{data burst} which is sent using
\textit{one-way resource reservation}. The burst is preceded in time
by a control packet, called \textit{Burst Header Packet} (BHP), which is sent on a separate control wavelength
and requests resource allocation at each switch. When the control
packet arrives at a switch, the capacity is reserved in the
cross-connect for the burst. If the needed capacity can be reserved at a given time, the burst
can then pass through the cross-connect without the need of buffering or
processing in the electronic domain.

	Since data bursts and control packets are sent out without waiting
for an acknowledgment, the burst could be dropped due to
resource contention or to insufficient offset time if the burst
catches up the control packet. Thus, it is clear that burst
contention resolution approaches play an essential role to reduce
the BLR in OBS networks \cite{contention2}.

	Burst contention can be resolved using several approaches, such as
\textit{wavelength conversion}, \textit{buffering} based on
\textit{fiber delay line} (FDL) or \textit{deflection routing}. Since deflection cannot eradicate the burst loss, retransmission at the OBS layer has been suggested by Torra et al. \cite{firstRetransmission} to avoid resource contention.
Another approach, called \textit{burst segmentation}, resolves
contention by dividing the contended burst into smaller parts called
\textit{segments}, so that a segment is dropped rather than the
entire burst. All these approaches are reactive and attempt to minimize the
BLR based on local information available at the core node. Whereas
a proactive approach that avoids burst losses before they occur is
desirable.

This paper introduces a novel algorithm called \textit{Graphical Probabilistic Routing Model} (GPRM) for OBS networks in order to build more effective routing tables and hence to reduce the BLR without affecting the end-to-end delay. To the best of our knowledge, our work is the first that proposes a graphical probabilistic model to the problem of optimal routing in OBS networks. Reinforcement Learning algorithms for path selection and wavelength selection have been proposed \cite{reinforcementOBS}. However predetermined routes are computed. GPRM does not need any precomputed paths since a path is constructed based on local knowledge on adjacent hops.

From the ingress node to the destination, the BHP selects at each intermediate node the next hop by using a lookup in the routing table. These routing tables are periodically updated by the learning process of the bayesian network. GPRM algorithm exploits the exchange of \textit{Positive Acknowledgement} (ACK) and 
\textit{Negative Acknowledgement} (NACK) messages in order to update bayesian networks so that each node learn the status of the other nodes. At each OBS node, an agent is placed and makes dynamic updates to his routing table by using a bayesian network. Our
approach allows us to update the local policies while
avoiding the need for a centralized control or a global knowledge of
the network state.

We choose Bayesian networks \cite{Bayesian} for our learning models
because of their expressiveness and more elegant graphical
representation compared to other black box machine
learning models. Bayesian networks, sometimes called belief
networks or graphical models, can be designed and interpreted
by domain experts because they explicitly communicate the
relevant variables and their interrelationships.

This paper is organized as follows. Section \ref{sec:motivation} gives the motivation of this work and in section \ref{sec:proposedModel}, we present the proposed model. Finally,
Section \ref{sec:SimulationTestbedAndResults} shows simulation testbed and results and Section \ref{sec:ConclusionAndFutureWork} contains the conclusion and future work.
%
\section{Motivation}
\label{sec:motivation}

	The traffic between two cities highly depends on the population, the number of employees and on the number of hosts \cite{article:refTransportScenarios}. Thus, the reality is that the traffic is not uniformally distributed. So intuitively static algorithms such as the shortest path are not effective in terms of network utilization.

	The main idea of this work is to propose an \textit{intelligent} routing mechanism for OBS Networks to adapt routing paths according to the network environment (traffic variations, link or node failure, topology changes). From a machine learning
perspective, it is desirable for the routing mechanism to tune itself into a
systematic, mathematically-principled way.

A probabilistic graphical model specifies a family of probability distributions which can be represented in terms of a graph \cite{livre:rbRef}. A bayesian network is a probabilistic graphical model and a directed acyclic graph. Nodes represent variables and links represent dependencies.

\begin{figure}
\centering
\includegraphics[scale=0.45]{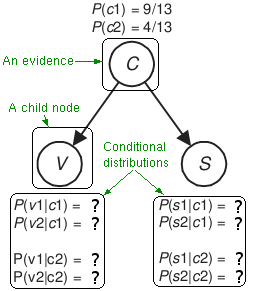}
\caption{A bayesian network example}
\label{fig:bayesianNetworkExample}
\end{figure}

	For example Fig. \ref{fig:bayesianNetworkExample} presents a very simple example of a bayesian network composed of 3 nodes (the variables) and 2 directed links (the dependencies). The main components of a variable are his state and a conditional distribution (a conditional probability table). A state is typically a possible category of values. For example let's say we have a variable \textit{Temperature} which can be either \textit{Raining} ($Ra$), \textit{Cloudy} ($Cl$) or \textit{Sunny} ($Su$). In this case, possible states are \{$Ra$, $Cl$, $Su$\}. In the example (Fig. \ref{fig:bayesianNetworkExample}), variable $C$ is called an evidence since it is a known information. Variables $V$ and $S$ depend on the state of the variable $C$. Variable $C$ is a parent variable of $V$ and $S$. In general we say that a variable depends on his connected parents. The conditional distribution of a given variable gives probabilities for all combinations of the given variable's states and all his parents states. More formally, the joint distribution of a bayesian network which contains a set of $N$ variables ${X_1, X_2, ..., X_N}$ is given by the product of each node distribution and its parents :

\begin{equation}
\label{formula:jointDistributions} 
P(X_1, X_2, ..., X_N) =  \prod_{i = 1}^{N}P(X_i | parents(X_i))
\end{equation}
	
	If a variable has no parent, it is said to be unconditional. Another important concept is the notion of inference. Resolving a probabilistic inference is done by applying Bayes' theorem \cite{livre:rbRef}.  If a given node $X_i$ has $M$ evidences, the probability that variable $X_i$ is in the state $x_i$ is found by doing the following inference:

\begin{equation}
\label{formula:inference} 
P(X_i = x_i | E_1 = e_1, E_2 = e_2, ..., E_M = e_M)
\end{equation}

	This inference application is trivial. Several more advanced applications can be done \cite{livre:rbRef}. 

A bayesian network is a model to represent the knowledge as well as 
a conditional probability calculator. It is also a learning model that can be used to learn the parameters (conditional distributions) or the structure of the bayesian network. The proposed model presented in Section \ref{sec:proposedModel} uses parameter learning to optimize the routing table. The use of a bayesian network, in our study, is motivated by the followings:

\begin{itemize}
\item The probabilistic formulation of a bayesian network is exploited for performing the resource reservation. Next hop wavelength selection is then represented by a conditional probability calculated using our model. 

\item The \textit{best} next hop must be selected based on several metrics. Those metrics are represented by evidences in our model. A bayesian network inference is used to calculate the probability to reserve the bandwidth based on several evidences (metrics). For example, some evidences in OBS networks are: BLR, end-to-end delay, load, offset time, etc.

\item The concept of linking variables by using arrows reduces the complexity of interrelationships between relevant variables. Thus, the studied system can be easily modelized and modified graphically by the domain expert. 

\item The learning capability of the bayesian network allows the network to systematically be adapted to its environment (topology changes, traffic variation, node and link failure). The probabilistic graphical model can recompute and reoptimize automatically routing tables.
\end{itemize}

%
\section{Proposed model}
\label{sec:proposedModel}

	A detailed description of the proposed model is given in this section as well as the routing table. Then, the signaling scheme and the notification packets are defined.

\subsection{GPRM description}

\begin{figure}
\centering
\includegraphics[scale=0.5]{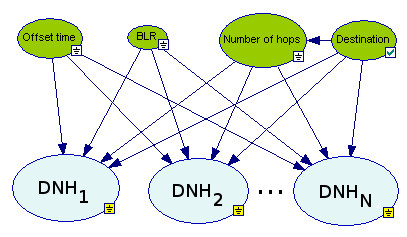}
\caption{The Graphical Probabilistic Routing Model (GPRM) - A Bayesian Network}
\label{fig:gprmModel}
\end{figure}

\begin{figure}
\centering
\includegraphics[scale=0.45]{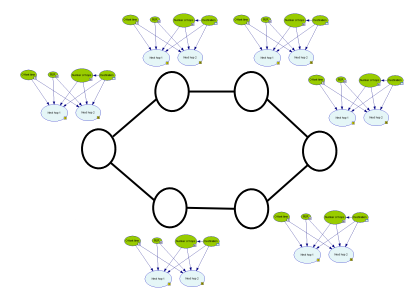}
\caption{A very basic topology with GPRM}
\label{fig:veryBasicTopologyWithGPRM}
\end{figure}

\begin{figure}
\centering
\includegraphics[scale=0.35]{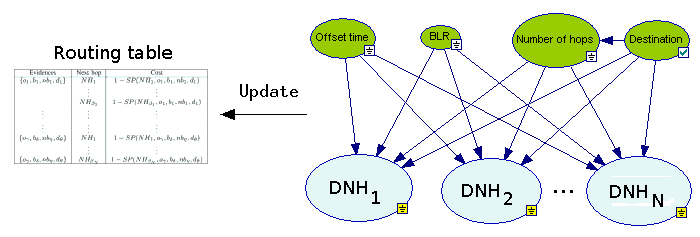}
\caption{GPRM updating the routing table}
\label{fig:gprmLinkedToRoutingTable}
\end{figure}

	The proposed model is a bayesian network composed of known information (evidences) and decision nodes (Fig. \ref{fig:gprmModel}). There is one bayesian network for each network node within the OBS topology (Fig. \ref{fig:veryBasicTopologyWithGPRM}). The main functionality of GPRM is the selection of the next hop for the forwarding process. Obviously, the forwarding process must be fast and a typical fast routing table lookup is computed periodically by the proposed model (Fig. \ref{fig:gprmLinkedToRoutingTable}). This routing table is used when the BHP attempts (in the electronic domain) to reserve the resource for the data burst.

	GPRM is composed of four evidences and one decision node for each possible next hop. A lookup to the routing table is done according to evidences in order to successively get the best next hop in terms of probability of success to reach the destination. GPRM includes the following evidences:

\begin{figure}
\centering
\includegraphics[scale=0.35]{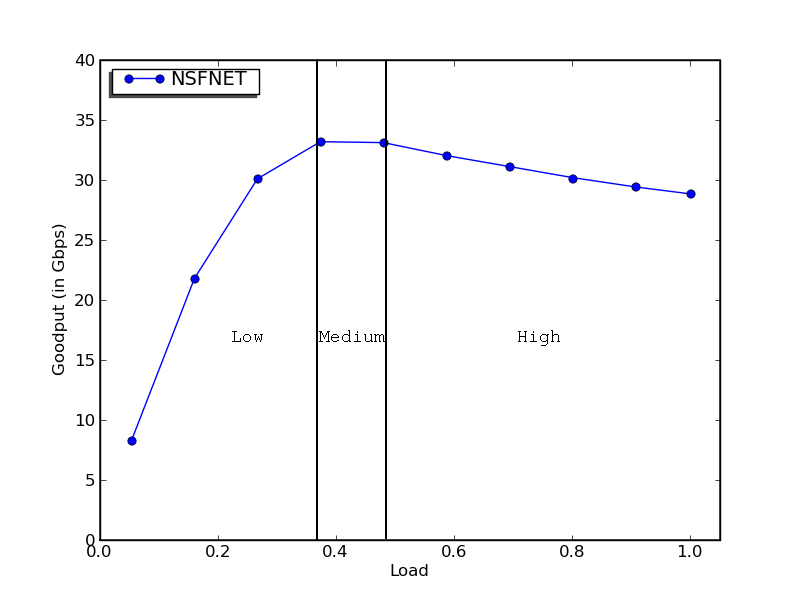}
\caption{Low, medium and high traffic delimitations by using the goodput}
\label{fig:delimitationLowMedHigh}
\end{figure}

\begin{itemize}
\item Offset time ($O$): The offset time has a significant impact on burst loss since if the offset time is insufficient, the burst will be dropped. Consequently the offset time is a relevant metric to be considered in order to select the next hop. The states of this evidence are based on the number of hops to reach the destination from the source (we used 0 to 15). Knowing the path length, the offset time can be easily calculated. 
\item BLR ($B$): The BLR is used to categorize statistics. In our study, we consider three possible states for this variable :  \textit{Low}, \textit{Medium} and \textit{High} (Fig. \ref{fig:delimitationLowMedHigh}).
\item Number of hops ($NB$): The states of this variable depends on the destination. However it has been added as an input variable for decision nodes.
\item Destination ($D$): Possible states of this variable are the OBS node identifiers.
\end{itemize}

	GPRM also includes one decision node per possible next hop. Each decision node has two possible states: \textit{Success} (noted $\oplus$) and \textit{Failure} (noted $\ominus$).

	Let $k$ be an OBS node identifier, $DNH_k$ expresses the bayesian decision node of the OBS next hop $k$, the joint probability function of the proposed model is given by:

\begin{align} 
\label{formula:jointProb}  
P(O, B, NB,& D, DNH_1, ..., DNH_N) = \nonumber \\
 & \left ( \prod_{i = 1}^{N} P(DNH_i|O, B, NB, D) \right ) * \nonumber \\
 & P(O) * P(B) * P(NB | D) * P(D) 
\end{align}

	The maximum a posteriori of $DNH_k$ is defined by:

\begin{equation}
\label{formula:inference1}  
MAP_{DNH_k} = \mathop{{\arg\max}\vphantom{\sim}}\limits_{\displaystyle _{\mathbf \varphi}} P(\varphi | o, b, nb, d)
\end{equation}
where $\varphi \in \{\oplus, \ominus\}$ which is a possible value of node $DNH_k$ and where $o, b, nb, d$ are possible values of the evidences in the bayesian network ($o$ is a possible state of the \textit{Offset time} variable, etc.).

\begin{equation}
\label{formula:inference2}  
MAP_{DNH_k} = \mathop{{\arg\max}\vphantom{\sim}}\limits_{\displaystyle _{\mathbf \varphi}} \frac{P(\varphi) P(o, b, nb, d|\varphi)}{P(o, b, nb, d)}
\end{equation}

\begin{equation}
\label{formula:inference2}  
MAP_{DNH_k} = \mathop{{\arg\max}\vphantom{\sim}}\limits_{\displaystyle _{\mathbf \varphi}} P(\varphi) P(o, b, nb, d|\varphi)
\end{equation}

	If we assume that the evidences are independent, the maximum a posteriori can be approximated as follows:

\begin{equation}
\label{formula:inference2}  
MAP_{DNH_k} \approx \mathop{{\arg\max}\vphantom{\sim}}\limits_{\displaystyle _{\mathbf \varphi}} P(\varphi) P(o|\varphi) P(b|\varphi) P(nb|\varphi) P(d|\varphi)
\end{equation}

	Let $SP(DNH_k, o, b, nb, d)$ (Success Probability, $DNH_k = \oplus$) be the approximation of $MAP_{DNH_k}$ when $DNH_k = \oplus$. 

\subsection{Routing table}

	The proposed model uses a different routing table compared to the typival approaches such as the shortest path in order to consider GPRM's bayesian network. In most routing approaches, metrics used in the routing table are $<Destination, Next\ hop, Cost>$. However GPRM's routing table uses $<Evidences, Next\ hop, Cost>$ where evidences add granularity in order to route the traffic more effectively. GPRM's routing table is a fast routing table defined as follows:

\begin{table}[h]
\begin{center}
\caption{GPRM routing table}
\label{tableRoutingTable}

\begin{tabular}{|c|c|c|}

\hline
\centering
  Evidences & Next hop & Cost \\
  \hline
  $\{o_1, b_1, nb_1, d_1\}$ & $NH_{1}^{1}$ & $1 - SP(DNH_{NH_{1}^{1}}, o_1, b_1, nb_1, d_1)$ \\
   & \vdots & \vdots \\
   & $NH_{\beta_1}^{1}$ & $1 - SP(DNH_{NH_{\beta_1}^{1}}, o_1, b_1, nb_1, d_1)$ \\
	\hline
  \vdots & \vdots & \vdots \\
  \vdots & \vdots & \vdots \\
	\hline
  $\{o_\gamma, b_\delta, nb_\eta, d_\theta\}$ & $NH_{1}^{EP}$ & $1 - SP(DNH_{NH_{1}^{EP}}, o_\gamma, b_\delta, nb_\eta, d_\theta)$ \\
   & \vdots & \vdots \\
   & $NH_{\beta_{EP}}^{EP}$ & $1 - SP(DNH_{NH_{\beta_{EP}}^{EP}}, o_\gamma, b_\delta, nb_\eta, d_\theta)$ \\
  
  \hline

\end{tabular}

\end{center}

\end{table}

where $\beta_i$ expresses the number of different next hops depending on the evidences of the $i$ permutation, $\gamma, \delta, \eta, \theta$ are the number of possible states for evidence variables. $NH_{i}^{j}$ represents the $i^{th}$ next hop of the $j^{th}$ evidence permutation. $NH_{1}^{j}$ represents the minimum cost of the $j^{th}$ evidence permutation. For example if we have the evidence permutation $\{o_1, b_1, nb_1, d_1\}$ and $\beta_1 = 3$, we could have the following next hops: $\{1, 3, 7\}$. The column \textit{Evidences} represents the combinations of all states of all evidences. The number of evidence permutations is defined by:

\begin{equation}
\label{formula:permutations}  
EP = \gamma * \delta * \eta * \theta
\end{equation}

Thus, the number of rows in the routing table is defined by:

\begin{equation}
\label{formula:nbLignesTable}  
N_{Rows} = \sum_{i=1}^{EP} \beta_i
\end{equation}

	The cost is expressed by:

\begin{equation}
\label{formula:coutTableRoutage}  
Cost(NH, o, b, nb, d) = 1 - SP(NH, o, b, nb, d)
\end{equation}

	We note that for a given evidence permutation, next hops are sorted as follows:


\begin{equation}
\label{formulaSortRoutingTable} 
	\forall_{i=1}^{N} \forall_{j=1}^{\beta_i - 1} Cost(NH_{j}, o, b, nb, d) \le Cost(NH_{j+1}, o, b, nb, d)
\end{equation}

	The following algorithm defines the lookup mechanism to get the \textit{best} next hop:

\begin{algorithm}[H]
\SetLine
\linesnumbered

\KwData{$N$, a set of all OBS node identifiers.}

	\textit{At each node $i \in N$}

\ForEach{BHP} {
	\textbf{Extract evidences} \{o, b, nb, d\}

	\textbf{Find} the corresponding row ($k$) in the routing table.

	\textbf{Select} $NH_{k}^{1}$.
}

\caption{Look up in the routing for the best next hop according to evidences.}
\label{alg:orderNextHops}
\end{algorithm}

	The following algorithm maps the bayesian network to a fast routing table periodically:

\begin{algorithm}[H]
\SetLine
\linesnumbered

\KwData{$N$, a set of all OBS node identifiers.}
\KwData{$T$, time interval update.}

\textit{At each node $i \in N$}

\textit{At every $T$}

\ForEach{Evidence permutation $j^{th}$ $\{o, b, nb, d\}$} {

	\textbf{Locate} $\{o, b, nb, d\}$ in the routing table.
	
	\textbf{Get} next hop identifiers ($ids$) according to $\{o, b, nb, d\}$.

	$index \leftarrow []$

	\ForEach{$id \in ids$} {
		\textbf{Add} $(id, 1- SP(DNH_{id}, o, b, nb, d))$ by order of cost in $index$.
	}

	\textbf{Associate} $\{o, b, nb, d\}$ to $index$ in the routing table.
}

\caption{Maps the bayesian network to a fast routing table periodically.}
\label{alg:orderNextHops}
\end{algorithm}

\subsection{Signaling scheme and notification packets}

\begin{figure}
\centering
\includegraphics[scale=0.35]{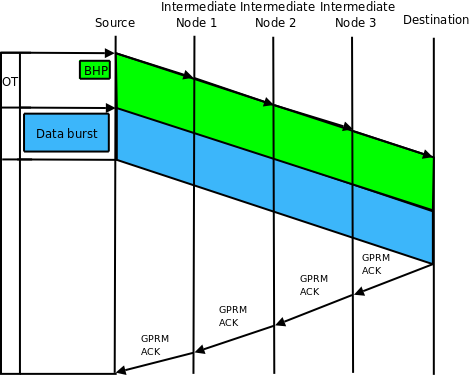}
\caption{Signaling scheme without contention}
\label{fig:supports_signaling_without_contention}
\end{figure}

\begin{figure}
\centering
\includegraphics[scale=0.35]{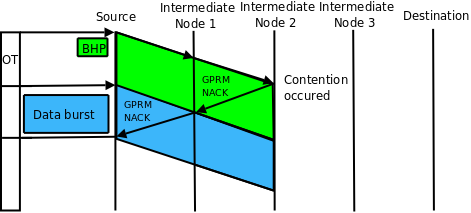}
\caption{Signaling scheme with contention}
\label{fig:supports_signaling_with_contention}
\end{figure}

	The proposed model uses the well-known JET signaling scheme \cite{obsGeneral2}. However notification packets are used in order to update GPRM's bayesian network. A \textit{positive acknowledgement} (ACK) is sent when a BHP reaches the destination (Fig. \ref{fig:supports_signaling_without_contention}). A \textit{negative acknowledgement} (NACK) is sent when a BHP can not reserve the bandwidth for the data burst (Fig. \ref{fig:supports_signaling_with_contention}). We note that these notification packets can also be used in an OBS scheme where retransmission is available in order to free buffered bursts. Also, evidences are stored in BHP and in notification packets in order to update GPRM's bayesian network. When an OBS node receives a notification packet, GPRM's bayesian network is updated and the routing table is refreshed at the next update period.

	The following algorithm (Algorithm \ref{alg:orderNextHops}) describes the reception of a notification packet and the update of the bayesian network according to evidences.

\begin{algorithm}[H]
\SetLine
\linesnumbered

\KwData{$N$, a set of all OBS node identifiers.}

	\textit{At each node $i \in N$}

\ForEach{ACK/NACK} {
	\textbf{Extract evidences} \{o, b, nb, d\}

	\textbf{Extract the last hop} ($LH$).

	\textbf{Extract the notification packet type} ($NPT$).

	$SP \leftarrow SP(DNH_{LH}, o, b, nb, d)$ 

	$SP' \leftarrow \alpha SP + (1 - \alpha)A$ 

	Where $\alpha \in [0..1]$ and $A$ is given by:

	\[
  A = \left\{
          \begin{array}{ll}
            1 & \mathrm{if}\ NPT = ACK \\
            0 & \mathrm{if}\ NPT = NACK \\
          \end{array}
        \right.
\]

	\textbf{Update}  the bayesian node $DNH_{LH}$ such that $SP'$ is the new value in the conditional distribution according to evidences.
}

\caption{Reception of a notification packet and update of the bayesian network according to evidences.}
\label{alg:orderNextHops}
\end{algorithm}

%
\section{Simulation Results}
\label{sec:SimulationTestbedAndResults}

\begin{figure}
\centering
\includegraphics[scale=0.35]{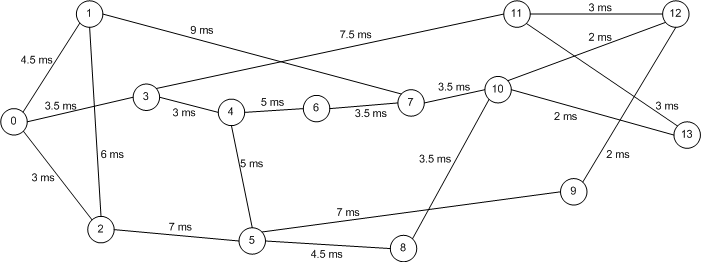}
\caption{NSFnet topology}
\label{fig:nsfnet}
\end{figure}

	Simulations are performed with NSFnet (Fig. \ref{fig:nsfnet}) topology by using \textit{Network Simulator 2} (ns-2) \cite{ns} with an extra module for OBS. The C++ library \textit{Structural Modeling, Inference, and Learning Engine} (SMILE) \cite{lien:smile} is used for the bayesian network. GPRM is compared to the well-known Shortest Path algorithm for the performance comparison. The shortest path algorithm always selects paths minimizing the number of hops. The proposed model (GPRM) defined in Section \ref{sec:proposedModel} is used, which tends to select paths in order to maximize link utilization and in order to decrease the BLR.

	The following simulation configuration is used:

\begin{itemize}
\item Each wavelength has 1 Gbit/s of bandwidth capacity.
\item Each link has 2 control channels and 4 data channels.
\item The mean burst size (noted $L$) equals 400 KB.
\item Packet and burst generation follows a Poisson distribution for the input packet rate and for the burst size.
\item Connections are distributed over the network proportionately to the well-known reference transport network scenario of the US Network \cite{article:refTransportScenarios}. 
\item Let $N$ be the number of nodes in the topology, $\xi_{i, j}$ the number of OBS connections between $i$ and $j$, $\lambda_{i, j, k}$ the number of bursts sent per second of the $k$ connection between $i$ and $j$, $\mu_{i}$ the capacity available at $i$, the load is given by:
\begin{equation}
\label{formula:load} 
	Load = \sum_{i = 1}^{N} \sum_{j = 1}^{N}\sum_{k=1}^{\xi_{i, j}} \frac{\lambda_{i, j, k} * L}{\mu_{i}}
\end{equation}

\item No deflection, retransmission or other contention resolution strategies are used.

\end{itemize}

\subsection{GPRM learning NSFnet topology without initial routing information}

\begin{figure}
\centering
\includegraphics[scale=0.45]{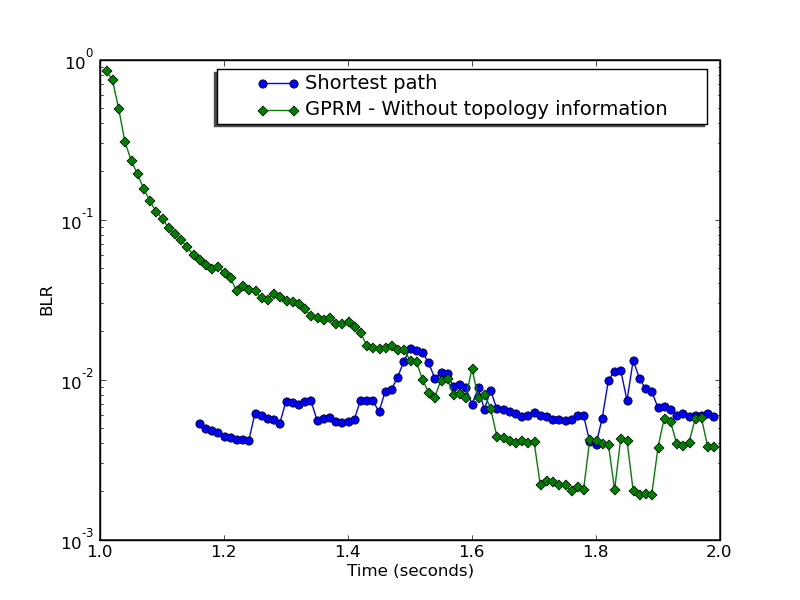}
\caption{GPRM learning NSFnet topology without initial routing information}
\label{fig:results_learning_topology}
\end{figure}

An implicit benefit of GPRM is the capability of an OBS node, without initial routing information, to learn his neighbors in order to route and distribute the traffic efficiently. It can be useful when faults happen in a topology since an automatic fault recovery mechanism is applied. The Shortest Path algorithm has initial information about the topology such as next hops, number of hops to reach destinations, etc. GPRM requires less than 1 second to learn how to distribute the traffic at least as effectively as the Shortest Path (Fig. \ref{fig:results_learning_topology}).

\subsection{Comparison of GPRM and Shortest path}

\begin{figure}
\centering
\includegraphics[scale=0.45]{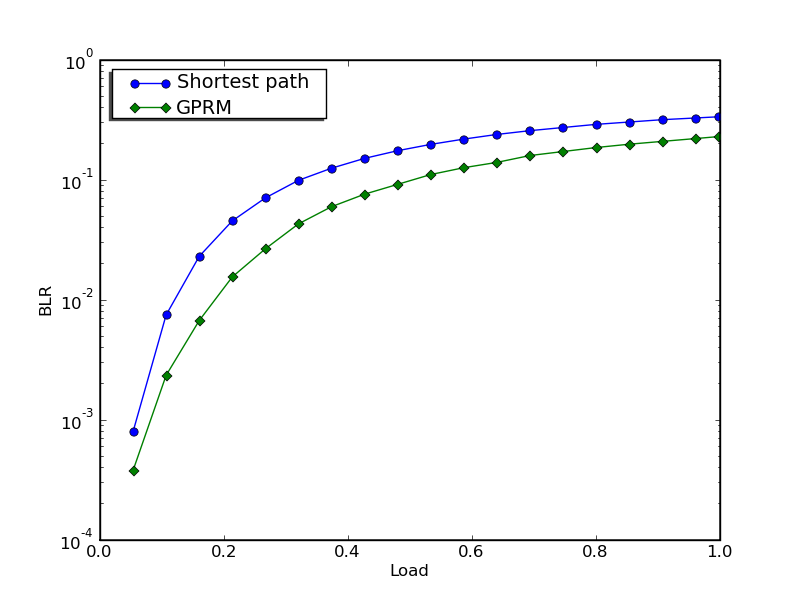}
\caption{BLR}
\label{fig:results_nsf_blr}
\end{figure}

	For the rest of the comparison, we assume that the initial routing information are available for both algorithms.

	GPRM gives significative improvements in terms of BLR even at high loads (Fig. \ref{fig:results_nsf_blr}). Let $N$ be the number of simulations where each simulation has a different load. The BLR gain is given by:

\begin{equation}
\label{formula:BLRGain} 
	BLR_{Gain} = \sum_{i=1}^{N} \frac{BLR_{SP, i} - BLR_{GPRM, i}}{BLR_{SP, i}}
\end{equation}

where $BLR_{SP, i}$ defines the BLR of the $i$ simulation by using the shortest path. The utilization gain is then defined by:

\begin{equation}
\label{formula:UtilizationGain}
	U_{Gain} = \sum_{i=1}^{N} \frac{U_{GPRM, i} - U_{SP, i}}{U_{SP, i}}
\end{equation}

\begin{figure}
\centering
\includegraphics[scale=0.45]{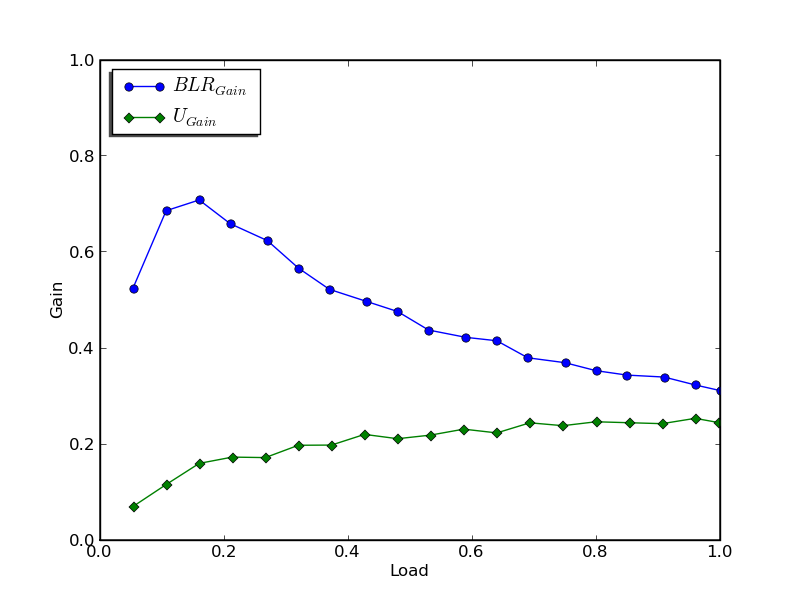}
\caption{BLR and utilization gains}
\label{fig:resultsGains}
\end{figure}

	We can observe that when the load is less than 0.5, using GPRM reduces the number of bursts dropped by 50 \% ($BLR_{Gain}$) compared to the shortest path algorithm (Fig. \ref{fig:resultsGains}).

\begin{figure}
\centering
\includegraphics[scale=0.45]{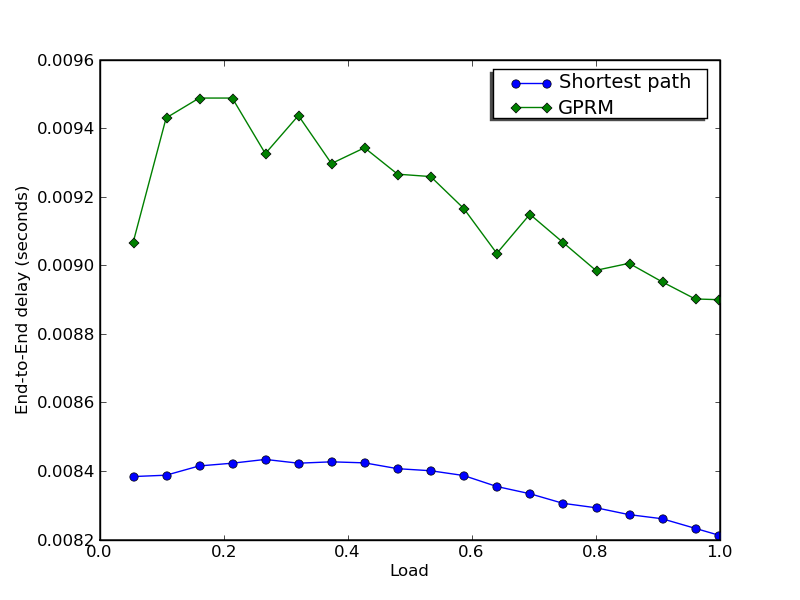}
\caption{End-to-End delay}
\label{fig:results_nsf_end2end}
\end{figure}

	GPRM gives at most 1 ms more of end-to-end delay compared to the shortest path algorithm (Fig. \ref{fig:results_nsf_end2end}). It can be explained by the fact that GPRM does not necessarily use the shortest path. GPRM can select next hop which requires more hops to reach the destination in order to use less utilized links. However, 1 ms does not have an impact on transport protocols such as TCP.

\begin{figure}
\centering
\includegraphics[scale=0.45]{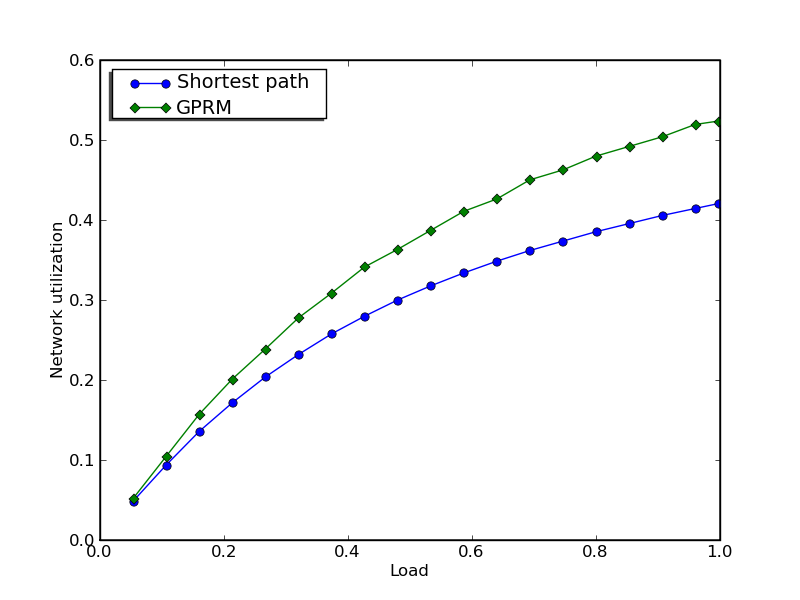}
\caption{Network utilization}
\label{fig:results_nsf_utilisation}
\end{figure}

	GPRM gives significative improvements in terms of network utilization (about 20 \% of $U_{Gain}$) as shown in Fig. \ref{fig:results_nsf_utilisation}. We can observe that a gain of 20 \% of network utilization can decrease more than 50 \% of BLR so the routing mechanism is very important in paradigms such as OBS where contention is an important issue.

%
\section{Conclusion and Future Work}
\label{sec:ConclusionAndFutureWork}

	This paper presents a novel routing scheme called \textit{Graphical Probabilistic Routing Model} (GPRM) that selects less utilized links by using a bayesian network. Decisions are based on local knowledge from a bayesian network updated each time an OBS node receives a notification packet. The bayesian model contains one decision node for each possible next hop from the current node. Conditional distributions are constructed from evidences, from known metrics. A routing table is periodically updated by using the bayesian network in order to not penalize the forwarding process. Permutations of all states of all evidences are included in the routing table in order to make decisions more effective. GPRM is capable to learn an unknown topology as well as recover network faults. Simulation results show that the proposed model performs efficiently in NSFnet since it decreaces significantly the BLR and increases the network utilization without affecting considerably the end-to-end delay.

A possible future step of this research is to combine several contention resolution strategies in a dynamic way because we believe that the feasibility of OBS requires effective and adaptive algorithms to overcome the burst loss issue.

\bibliographystyle{IEEE}
\bibliography{GPRM}

\end{document}